\documentclass{amsart}

\newcommand{\be}{\begin{equation}}
\newcommand{\ee}{\end{equation}}
\newcommand{\bea}{\begin{array}}
\newcommand{\ea}{\end{array}}
\newcommand{\beqa}{\begin{eqnarray}}
\newcommand{\eeqa}{\end{eqnarray}}
\newcommand{\bean}{\begin{eqnarray*}}
\newcommand{\eean}{\end{eqnarray*}}

\def\up#1{\leavevmode \raise.16ex\hbox{#1}}

\newcommand{\gapproxeq}{\lower
 .7ex\hbox{$\;\stackrel{\textstyle >}{\sim}\;$}}
\newcommand{\lapproxeq}{\lower .7ex\hbox{$\;\stackrel
{\textstyle <}{\sim}\;$}}

\newcounter{appendice}

\def\thebibliography#1{{\bf REFERENCES\markboth
 {REFERENCES}{REFERENCES}}\list
 {[\arabic{enumi}]}{\settowidth\labelwidth{[#1]}\leftmargin\labelwidth
 \advance\leftmargin\labelsep
 \usecounter{enumi}}
 \def\newblock{\hskip .11em plus .33em minus -.07em}
 \sloppy
 \sfcode`\.=1000\relax}

\begin{document}

\centerline{ \LARGE  Canonical Quantization, Space-Time Noncommutativity}
\centerline { \LARGE and Deformed Symmetries in Field Theory}
\vskip .5cm
\vskip 2cm
\centerline{ {\sc    Marcos Rosenbaum$^{a}$, J. David Vergara$^{b}$
and L. Roman Juarez$^{c}$} }
\vskip 1cm
\begin{center}
Instituto de Ciencias Nucleares,\\
Universidad Nacional Aut\'onoma de M\'exico,\\
 A. Postal 70-543 , M\'exico D.F., M\'exico\\
{\it a)mrosen@nucleares.unam.mx,\\
b)vergara@nucleares.unam.mx,\\
c)roman.juarez@nucleares.unam.mx}%
\end{center}
\vskip 2cm \vspace*{5mm} \normalsize \centerline{\bf ABSTRACT}
Within the spirit of Dirac's canonical quantization, noncommutative
spacetime field theories are introduced by making use of the
reparametrization invariance of the action and of an arbitrary
non-canonical symplectic structure. This construction implies that
the constraints need to be deformed, resulting in an automatic
Drinfeld twisting of the generators of the symmetries associated
with the reparametrized theory. We illustrate our procedure for the
case of a scalar field in  1+1- spacetime dimensions, but it can be
readily generalized to arbitrary dimensions and arbitrary types of
fields. \vspace{2cm}
\newpage

\section{Introduction}
 It has been considered as common wisdom among practitioners of
noncommutative field theory that at the first quantization level,
fields are elements of an algebra where multiplication is deformed
by means of the Moyal $\star$-product \cite{szabo}. This anzatz,
which originated in a basically heuristic fashion from some results
in string theory \cite{Seiberg}, is based on an analogy with the
Weyl-Wigner-Groenewold-Moyal (WWGM) formalism of Quantum Mechanics.
But in Quantum Mechanics time is a parameter of the theory and, in
order for spacetime to have a truly  noncommutativity physical
meaning we need to consider both space and time as observables
represented by noncommutative operators and include them as dynamical
variables in an extended Heisenberg algebra.\\
Moreover, as we have shown elsewhere \cite{ros}, the $\star$-product
deformation of functions of spacetime then results naturally in the WWGM formalism when
considering in this extended context the algebra of the Weyl-equivalent functions corresponding
to operator functions of the Heisenberg space and time operators.

Other approaches for constructing a noncommutative spacetime Quantum
Mechanics have been based on the idea of promoting the time
parameter to the rank of a coordinate by means of a
reparametrization, whereby time becomes a function $t(\tau)$ of a
new parameter $\tau$ and thus becomes a coordinate on the same level
as the spatial coordinates $x^i(\tau)$, either by fixing the gauge
degrees of freedom  \cite{Pinzul}, \cite{Banerjee}, \cite{Ghosh} or
by deforming the symplectic structure of the theory \cite{Ros1}.

 An important feature of these formulations is that,
because additional degrees of freedom are added to the original theory,
first class constraints appear in the reparametrized theory. In order
to eliminate these additional degrees of freedom one can apply
gauge conditions or follow Dirac's quantization method and operate
with the constraints on the state vectors in order to obtain the physical
states of the system.

Now, when going on to field theory both the time and space coordinates
play the role of parameters of the field, so applying commutation relations
to them is, to say the least, even more unclear; as it is the relation of this
procedure to the operator spacetime noncommutativity in Quantum Mechanics,
particularly when we view the latter as a minisuperspace of the former and in
the light of what we have just said above.

In order to shed some additional insight on some of these issues,
we explore in the present work how the above refereed reparametrization
formalism can
be extended to the case of field theory on a noncommutative
space-time. However, since we are now dealing with a system with an
infinite number of degrees  of freedom, the basic idea here is to
promote the coordinates of the space-time, that are the parameters
on which the field depends, to new fields in the ensuing
reparametrized theory. This idea in not new in the case of commutative spacetime.
 For example in
\cite{karel} such a construction of a field theory was used as a model when considering
the canonical quantization of gravity. Making use of the results in that work,
it is possible to construct the reparametrized theory for any field theory, with
as many constraints as the number of coordinate fields being added.
In addition, as it occurs in the case of General Relativity,
the parametrized field theory is also invariant under diffeomorphisms, so such a construction
provides an ideal arena for studying these symmetries at the quantum level there.
It is interesting to note that this idea was also used in
the context of string theory as a means for constructing
a theory which would be independent of the background
\cite{Das}. \\
Once the spacetime coordinates are promoted to the rank of fields,
it does make sense to impose commutation relations among them. This
can be achieved by deforming the symplectic structure in the original
theory and thus arriving at a noncommutative field theory. Such a theory
is already at the first quantization level radically different
from the usual one, because - since the coordinate fields do not commute -
we can not use their eigenstates as configuration space
bases to construct amplitudes of the state vector, which will then
necessarily have to be either functions of both the eigenvalues of the momenta field operators
as well as of some of the coordinate fields (those that commute among themselves), or only of
the eigenvalues of the commuting momenta fields.

Another important point that we analyze in this paper is the
deformed symmetries that appear in the noncommutative theory.
According to our procedure, the nature of these deformed symmetries
appears automatically since, when deforming the symplectic structure
the algebra of the constraints is broken and, in order to preserve
it, it is necessary to deform the generators of the symmetry by
means of what turns out to be a Drinfeld twist. The algorithm
suggested by our procedure for this twist is quite straightforward
to implement and can be readily generalized to other types of
$\star$-products as well as to situations where noncommutativity
involves both spacetime and momenta variables.

\section{Spacetime Noncommutativity in Field Theory}

In a previous paper \cite{Ros1} noncommutative space-time quantum
mechanical theories were constructed by using a reparametrization
invariant action where the time parameter is elevated to the rank of
a dynamical variable. Furthermore, in order to consider the
noncommutativity between the space-time coordinates, an arbitrary
non-canonical symplectic structure was introduced that, together
with Dirac's Hamiltonian method, leads to Dirac brackets for the
space-time dynamical variables, which when quantized may be
interpreted as noncommutative. As mentioned in the Introduction, we shall
apply this procedure to the case of fields in order to investigate the
implications of noncommutativity  of spacetime as field variables on
the algebra of the reparametrized fields.
\subsection{Reparametrization of the scalar field}
To illustrate the procedure, consider for
simplicity the case of a scalar field in a $D+1-$dimensional
Minkowski spacetime with signature $(1,-1,\dots ,-1)$ and with a
potential $V(\phi)$. The corresponding action is then
\begin{equation}\label{scalarf}
S = \int dxdt \left( \frac{1}{2} \eta ^{\mu \nu } \partial _{\mu}  \phi
\partial _{\nu}  \phi -V(\phi)\right) .
\end{equation}
In order to parameterize the full spacetime, let us write
\begin{equation}\label{repara1}
\begin{gathered}
  t =t(\tau ,\boldsymbol\sigma), \hfill \\
  x^i = x^i\left( {\tau ,\boldsymbol\sigma } \right), \hfill \\
\end{gathered}
\end{equation}
so that the new action in terms of the new parameters $\tau,
\boldsymbol\sigma$ reads
\begin{equation}\label{ftact}
S = \int {d\tau d^D\sigma {\sqrt {-g} }}\left(\frac{1} {2} g^{\mu
\nu }\partial _\mu \phi
\partial _\nu  \phi -V(\phi)\right),
\end{equation}
with the inverse metric  $g^{\alpha\beta}$ given by
\begin{equation}\label{invmetrd}
g^{\alpha\beta}= \frac{\partial\sigma^\alpha}{\partial x^\mu}
\frac{\partial \sigma^\beta}{\partial x_\mu}
\end{equation}
and $g:=\det(g_{\mu\nu})$ where  $\sqrt{-g}=J$ is the Jacobian of
the transformation. Also, in (\ref{ftact}) we are making the
identification
$\partial_0 \equiv\partial_\tau$ and $\partial_i =\partial_{\sigma_i}$.\\
The canonical momentum associated to the field $\phi$ is
\begin{eqnarray}
P_\phi&=& J \frac{\partial \tau}{\partial x^\mu}\frac{\partial
\sigma^\alpha} {\partial x_\mu}\frac{\partial \phi}{\partial
\sigma^\alpha} ,\;\;\;\; \sigma^0 =\tau,\; \sigma^i\equiv\sigma^i,
\label{kmomenta1},
\end{eqnarray}
and, following \cite{karel}, we define the canonical momenta
associated to the spacetime coordinates as
\begin{eqnarray}
p_\nu &\equiv& -J \frac{\partial \tau}{\partial x^\mu}
{T^{\mu}}_{\nu} \label{kmomenta2},
\end{eqnarray}
where ${T^{\mu}}_{\nu}= \partial^\mu \phi
\partial_\nu \phi-\delta^\mu_\nu (\frac{1}{2}\partial^\rho \phi
\partial_\rho \phi-V(\phi))$ is the unparametrized energy-momentum
tensor of the field. In terms of this momenta the Hamiltonian action
becomes
\begin{equation}\label{Hamilact1}
    S=\int d\tau d^D\sigma \left(P_\phi \dot \phi + p_\mu \dot x^\mu -
    \lambda^\nu\left(p_\nu + J \frac{\partial \tau}{\partial x^\mu} {T^{\mu}}_{\nu}\right) \right),
\end{equation}
where we have introduced the definition of the momenta
(\ref{kmomenta2}) as Hamiltonian constraints due to the fact
that the right hand side of (\ref{kmomenta2}) is independent
of the velocities when the energy-momentum tensor is expressed
as a function of the canonical variables $\phi, P_\phi$ \cite{karel}.\\

We can write an alternate expression for the action (\ref{Hamilact1}), based on the
ADM-type decomposition of spacetime
$\Sigma \times \Bbb R$, where $\Bbb R$ is the temporal direction and
$\Sigma$ is a space-like hypersurface of constant $\tau$,
by introducing the vectors ${\bf s}_i$ with components
$s^{\mu}_i =\partial_{\sigma^i}x^\mu $ tangent to $\Sigma$ and the unit
vector $\hat{\bold n}$, normal to this hypersurface, with components
\be\label{normal2} n^\mu = \left(\sqrt {g^{00}} \dot x^\mu +
\frac{g^{0i}}{\sqrt {g^{00}}} \frac{\partial x^\mu}{\partial
\sigma^i} \right), \;\;\;i=1,\dots,d. \ee
 Furthermore, constructing from $s^{\mu}_i$ the orthonormal
basis $\hat{\bf u}_i =\alpha_{i}^j{\bf s}_j$,
 we can write the (D+1)-vector constraint ${\boldsymbol \Pi}$,
with components $\Pi_\nu\equiv p_\nu +J \frac{\partial \tau}{\partial x^\mu}
{T^{\mu}}_{\nu}$, as
\be\label{projconst}
 {\boldsymbol \Pi}\equiv(\hat{\bold n}\hat{\bold n} +\gamma^{ij} \hat{\bold u}_i \hat{\bold u}_j)\cdot {\boldsymbol \Pi}=
\hat{\bold n} \Phi_0 +\gamma^{ij} \hat{\bold u}_i \Phi_j, \ee where
\be\label{dyad} {\bold I}:= (\hat{\bold n} \hat{\bold n}
+\gamma^{ij} \hat{\bold u}_{i} \hat{\bold u}_{j}), \ee is the unit
dyadic, multiplication is with the Lorentzian metric,
\be\label{superh2} \Phi_0 :=\hat{\bold n}\cdot {\boldsymbol \Pi}=
n^\mu (p_\mu + J \frac{\partial \tau}{\partial x^\nu}
{T^{\nu}}_{\mu} )=\frac{1}{2\sqrt{-\gamma}}\left( P_\phi ^2 +\gamma
\gamma^{ij}\partial_{\sigma^i}\phi \partial_{\sigma^j}\phi\right) +
n^\mu p_\mu +
 \sqrt{-\gamma}\; V\left( \phi  \right),
\ee
\be\label{superm2}
\Phi_j :={\bold s}_{j}\cdot {\boldsymbol \Pi}= (\partial_{\sigma^j} x^{\mu}) (p_\mu + J \frac{\partial
\tau}{\partial x^\nu} {T^{\nu}}_{\mu} ) =P_{\phi}
\partial_{\sigma^j}\phi +p_\mu \partial_{\sigma^j} x^\mu,
\ee and where  $\gamma_{ij} \equiv g_{ij}$ is the D-metric of the
$\Sigma$-hypersurface, $\gamma^{ij}$ is the inverse matrix to
$\gamma_{ij}$ and $\gamma$ is the determinant of $\gamma_{ij}$.
Inserting now (\ref{projconst}) into (\ref{Hamilact1}) we can write
\begin{equation}\label{Hamilact11}
    S=\int d\tau d^D\sigma \left(P_\phi \dot \phi + p_\mu \dot x^\mu -
    N{\mathcal H}_{\perp} -N^i {\mathcal H}_i \right),
\end{equation}
after identifying the proyections
$(-\gamma)^{-\frac{1}{2}}(\boldsymbol \lambda \cdot \hat{\bold
n})$,\; $\gamma^{jk} \alpha_{k}^i\boldsymbol \lambda \cdot
\hat{\bold u}_j$  of the Lagrange multipliers with the lapse and
shift functions $N$ and $N^i$, respectively, so that ${\mathcal
H}_{\perp}=\sqrt {-\gamma}\Phi_0$ is the super-Hamiltonian  and
\;${\mathcal H}_i =\Phi_i$ are the super-momenta for the system. \\
The Poisson brackets of these super- Hamiltonian and super-momenta
are given by \cite{Dirac}
\begin{eqnarray}\label{fcc4d}
\{{\mathcal H}_{\perp} (\boldsymbol\sigma,\tau) , {\mathcal H}_{\perp}
({\boldsymbol\sigma}',\tau\} &=& \sum_{i=1}^{D}({\mathcal H}_i
(\boldsymbol\sigma,\tau)
 +{\mathcal H}_i ({\boldsymbol\sigma}',\tau))\partial_{\sigma^i}\delta(\boldsymbol\sigma-
 {\boldsymbol\sigma}'),\nonumber\\
\{{\mathcal H}_i (\boldsymbol\sigma,\tau), {\mathcal H}_k
({\boldsymbol\sigma}',\tau)\} &=&
({\mathcal H}_k (\boldsymbol\sigma,\tau)\partial_{\sigma^i}
\delta(\boldsymbol\sigma-{\boldsymbol\sigma}')
+{\mathcal H}_i ({\boldsymbol\sigma}',\tau))\partial_{\sigma^k}
\delta(\boldsymbol\sigma-{\boldsymbol\sigma}'),\\
\{{\mathcal H}_{\perp} (\boldsymbol\sigma,\tau), {\mathcal H}_i
({\boldsymbol\sigma}',\tau)\} &=&({\mathcal H}_{\perp}
(\boldsymbol\sigma,\tau) +{\mathcal
H}_{\perp}({\boldsymbol\sigma}',\tau))\partial_{\sigma^i}
\delta(\boldsymbol\sigma-{\boldsymbol\sigma}'),\nonumber
\end{eqnarray}
from where we see that the constraints are first class.

Let us now further simplify the calculations and the basic steps
leading to a noncommutative field theory by consider first our
scalar field to be propagating in a flat space-time with Minkowskian
coordinates $(t, x)$ and signature $(1, -1)$.  In this case
\begin{equation}\label{invmetr}
g^{\mu \nu }  =g^{-1} \left( {\begin{array}{*{20}c}
   { t^{'2} -x^{'2}  } & { - ( t'\dot t -x'\dot x) }  \\
   { - ( t'\dot t -x'\dot x) }  & { \dot t^2 -\dot x^2   }  \\
\end{array} } \right),
\end{equation}
and
\begin{equation}\label{gee}
g:=\det(g_{\mu\nu})=-(\dot t x'-\dot x t')^2 ,
\end{equation}
where the primes denote partials with respect to $\sigma$ while the dots
are partials with respect to $\tau$.\\
Explicit expressions for the momenta canonical to $t, x$ and $\phi$ can be derived from
(\ref{kmomenta1})
and (\ref{kmomenta2}) or, even simpler, directly from (\ref{ftact}), (\ref{invmetr}) and
(\ref{gee}). They are given by:
\begin{eqnarray}\label{mom}
p_t &=&-\frac{1}{{\sqrt {-g}}} (\dot t \phi'^2 -t'\phi'\dot \phi)
-x'V(\phi) -\frac{x'}{2g}[(t'^2 -x'^2 )\dot \phi^2 -2(t'\dot t-x'\dot x )\phi'\dot \phi
+(\dot t^2 -\dot x^2 )\phi'^2],\nonumber\\
p_x&=&\frac{1}{{\sqrt {-g}}} (\dot x \phi'^2 -x'\phi'\dot \phi) + t'V(\phi)
 +\frac{t'}{2g}[(t'^2 -x'^2 )\dot \phi^2 -2(t'\dot t -x'\dot x )\phi'\dot \phi
+(\dot t^2 -\dot x^2 )\phi'^2],\\
P_\phi &=& -\frac{1}{{\sqrt {-g}}}[(t'^2 -x'^2 )\dot \phi -(t'\dot t -x'\dot x )\phi']\nonumber.
\end{eqnarray}
From these expressions it can be readily verified that
\begin{equation}\label{ham}
p_t \dot t +p_x \dot x +P_\phi \dot \phi = {\mathcal L}={\sqrt {-g} }\left(\frac{1} {2} g^{\mu \nu }
\partial _\mu \phi
\partial _\nu  \phi -V(\phi)\right).
\end{equation}
Furthermore, because we are introducing the fields $t(\tau,\sigma)$
and $x(\tau,\sigma)$ as new degrees of freedom, the theory must have
constraints in the Hamiltonian formalism. Specifically, since
instead of our two original phase space degrees of freedom we now
have six, we thus need four relations which we can get by two
primary first class constraints,
and two gauge conditions.\\
The primary constraints follow from specializing (\ref{superh2}) and
(\ref{superm2}) to the case $D=1$ and are explicitly given by
\begin{equation}\label{ftcons}
\begin{gathered}
  {\mathcal H}_\perp  =\frac{1}{2}\left( P_\phi ^2 +\phi'^2 \right) +  p_t x' + p_x t' +
  \left( x^{'2}  - t^{'2}  \right)V\left( \phi  \right)  \approx 0, \hfill \\
  {\mathcal H}_1  = p_x x' + p_t t' + P_\phi  \phi ' \approx 0. \hfill \\
\end{gathered}
\end{equation}
Defining
\begin{equation}\label{def}
{\mathcal H}_{\perp,1} [f]:= \int d\sigma f(\sigma) {\mathcal H}_{\perp, 1} (\sigma,\tau),
\end{equation}
it can then be shown that
\begin{eqnarray}\label{fcc}
\{ {\mathcal H}_\perp[f], {\mathcal H}_\perp [g]\} &=& {\mathcal H}_1 [fg'-gf'],\nonumber\\
\{{\mathcal H}_1 [f], {\mathcal H}_1 [g]\} &=& {\mathcal H}_1 [fg'-gf'],\\
\{{\mathcal H}_\perp [f], {\mathcal H}_1 [g]\} &=&{\mathcal
H}_\perp[fg'-gf'].\nonumber
\end{eqnarray}
Moreover, since the test functions $f$ and $g$ are arbitrary,
we can take the functional derivatives of (\ref{fcc}) relative
to them to arrive at
\begin{eqnarray}\label{fcc2}
\{ {\mathcal H}_\perp (\sigma,\tau) ,  {\mathcal H}_\perp (\sigma',\tau\} &=&({\mathcal H}_1 (\sigma,\tau)
+{\mathcal H}_1 (\sigma',\tau))\delta'(\sigma-\sigma'),\nonumber\\
\{{\mathcal H}_1 (\sigma,\tau), {\mathcal H}_1 (\sigma',\tau)\} &=& ({\mathcal H}_1(\sigma,\tau)
+{\mathcal H}_1 (\sigma',\tau))\delta'(\sigma-\sigma'),\\
\{ {\mathcal H}_\perp (\sigma,\tau), {\mathcal H}_1 (\sigma',\tau)\}
&=&({\mathcal H}_\perp (\sigma,\tau) +{\mathcal
H}_\perp(\sigma',\tau))\delta'(\sigma-\sigma'),\nonumber
\end{eqnarray}
where $\delta'(\sigma-\sigma'):=\partial_\sigma \delta(\sigma-\sigma')$, which reproduce
(\ref{fcc4d}) for the case $D=1$. Note that
these constraints close in the constant $\tau$ Poisson brackets according to the Virasoro
algebra without a central charge and they are first-class, as we already know. But first
class constraints
are generically associated with gauge invariance, which in this case
is the invariance of the action (\ref{ftact}) under two-dimensional
reparametrizations, with its generators satisfying the algebra
(\ref{fcc2}).

Moreover since $H= \int d\sigma (N  {\mathcal H}_\perp +N^1
{\mathcal H}_1)$ is the Hamiltonian of the theory, it clearly
follows that
\begin{equation}\label{compat}
\dot{\mathcal H}_{\perp,1} =\{{\mathcal H}_{\perp,1},H\}\approx 0,
\end{equation}
so the constraints are preserved by the ``time" $\tau$ evolution.

Next, in order to introduce space-time noncommutativity in the Dirac
quantization procedure for the above theory, we need to implement an
additional general symplectic structure into our formalism.

 \subsection{Symplectic structure}
For this purpose consider the following general first order action:
\begin{equation}\label{ghactft}
S = \int d\tau d\sigma \left( A_a (z){\dot z}^a  - N \tilde{\mathcal H}_{\perp}
  - N^{1} \tilde{\mathcal H}_{1} \right),
\end{equation}
with symplectic variables  $z^a  = \left( {t,x,\phi ,p_t,p_x ,P_\phi  } \right)$.
 Here $\tilde{\mathcal H}_{\perp}$ and $\tilde{\mathcal H}_{1}$ are weakly zero
 and appropriately modified first-class constraints to be specified below.
The six potentials $A_a$ play the role of momenta canonically
conjugate to the $z^a$. The action (\ref{ghactft}) allows us to
generate an arbitrary symplectic structure associated to the Poisson
brackets in the Hamiltonian formulation, but in order that it be
equivalent to the action (\ref{Hamilact11}) for $D=1$, we need six
additional second-class primary constraints (these, together with
the two first-class constraints and their corresponding two
compatibility conditions, give the relations needed to eliminate ten
of the twelve degrees of freedom in the $z^a$'s).

The additional second-class constraints follow by noting that the canonical
momenta conjugate to $z^a$ are given by
\begin{equation}\label{canmom}
\pi_{z_a} = \partial_{\dot z^a}\left( A_a (z)\dot z^a   - N \tilde{\mathcal H}_{\perp}
  - N^{1} \tilde{\mathcal H}_{1} \right)=A_a(z) ,
\end{equation}
and since they are independent of the velocities they lead to the
constraints
\begin{equation}\label{secclass}
\chi_a =\pi_{z_a} - A_a \approx 0.
\end{equation}
Hence the action of our constrained system is now given by
\begin{equation}\label{ghactft2}
S = \int d\tau d\sigma \left( A_a (z)\dot z^a  - {\mathcal H}_T \right),
\end{equation}
with
\begin{equation}\label{totham}
{\mathcal H}_T =  N \tilde{\mathcal H}_{\perp}
  +N^{1} \tilde{\mathcal H}_{1} +\mu^a \chi_a .
\end{equation}
Note that from (\ref{secclass}) we have
\begin{equation}\label{seccom}
\{\chi_a , \chi_b \} =\frac{\partial A_b}{\partial z_a}  -\frac{\partial A_a}{\partial z_b}:=\omega_{ab},
\end{equation}
so the constraints $\chi_a$ are indeed second-class (note that the Poisson brackets here are to be
evaluated in the extended phase-space $(z^a, \pi_a)$ ).\\

 Moreover, in order that the consistency conditions
\be\label{comp1}
\dot \chi_a =\{\chi_a , \int d\sigma {\mathcal H}_T \} = -N\frac{\partial\tilde{\mathcal H}_{\perp}} {\partial z^a}
-N^{1} \frac{\partial \tilde{\mathcal H}_{1}}{\partial z^a} +\mu^b \omega_{ab}\approx 0,
\ee
\be\label{comp2}
\dot{\tilde{\mathcal H}}_{\perp, 1} = \{\tilde{\mathcal H}_{\perp, 1}, \int d\sigma {\mathcal H}_T \}
=\mu^a \{\tilde{\mathcal H}_{\perp, 1},\int d\sigma \chi_a\}\approx 0,
\ee
be satisfied, we need, solving (\ref{comp1}) for $\mu^a$, that
\begin{equation}\label{comp3}
\mu^a =\omega^{ab}\left(N \frac{\partial \tilde{\mathcal H}_{\perp}}{\partial z^b}+
N^1 \frac{\partial \tilde{\mathcal H}_{1}}{\partial z^b}\right),
\end{equation}
and also that
\begin{equation}\label{comp4}
\omega^{ab}\left( \frac{\partial \tilde{\mathcal H}_{\perp}}{\partial z^a}\frac{\partial \tilde{\mathcal H}_{1}}{\partial z^b}\right)\approx 0,
\end{equation}
which results from inserting (\ref{comp3}) into (\ref{comp2}) and using the arbitrariness
of the Lagrange multipliers.\\
Introducing now the Dirac brackets
\begin{equation}\label{dirac}
\{\xi ,\rho\}^\ast :=\{\xi ,\rho\} -\{\xi, \chi_a\}\omega^{ab} \{\chi_b ,\rho\},
\end{equation}
it readily follows that
\begin{equation}\label{dcomp}
\{\tilde{\mathcal H}_{\perp},\tilde{\mathcal H}_{1}\}^\ast =
\omega^{ab}\left( \frac{\partial \tilde{\mathcal H}_{\perp}}{\partial z^a}\frac{\partial \tilde{\mathcal H}_{1}}{\partial z^b}\right).
\end{equation}
Hence, in order to satisfy the compatibility condition (\ref{comp2})
we need to chose our modified constraints $\tilde{\mathcal
H}_{\perp}, \tilde{\mathcal H}_{1} $ such that their Dirac bracket
is weakly zero. We shall defer the proof that such a choice indeed
exist for later on, and note at this point that

\begin{equation}
\{\chi_a , \chi_b \}^\ast =0.
\end{equation}
We can therefore treat the $\chi_a$ as strongly zero in our
formalism, after replacing the Poisson brackets by the Dirac brackets.
Note also that (\ref{dirac}) implies
\begin{equation}\label{comm}
\{z^a , z^b \}^\ast=\omega^{ab},
\end{equation}
and by assuming further that the symplectic structure is determined by
\begin{equation}\label{symft}
\omega _{ab}  = \left( {\begin{array}{*{20}c}
   0 & 0 & 0 & { - 1} & 0 & 0  \\
   0 & 0 & 0 & 0 & { - 1} & 0  \\
   0 & 0 & 0 & 0 & 0 & { - 1}  \\
   1 & 0 & 0 & 0 & \theta  & 0  \\
   0 & 1 & 0 & { - \theta } & 0 & 0  \\
   0 & 0 & 1 & 0 & 0 & 0  \\
\end{array} } \right),\;\;
\omega ^{ab}  = \left( {\begin{array}{*{20}c}
   0 & \theta & 0 &  1 & 0 & 0  \\
   {-\theta} & 0 & 0 & 0 & 1 & 0  \\
   0 & 0 & 0 & 0 & 0 & 1  \\
   {-1} & 0 & 0 & 0 & 0  & 0  \\
   0 &{- 1} & 0 & 0 & 0 & 0  \\
   0 & 0 &{- 1} & 0 & 0 & 0  \\
\end{array} } \right),
\end{equation}
we find that (\ref{symft}), incorporates spacetime noncommutativity
into the formalism.
In particular upon quantization, the strong equations
$\chi_a =0$ need to be promoted to a relation between quantum operators:
\begin{equation}\label{qconst}
\hat \pi_{z_a} - \hat A_a =0,
\end{equation}
and we have from (\ref{comm}) that
at equal $\tau$
\begin{equation}\label{stcom}
\begin{gathered}
\left[ {\hat t\left( {\tau ,\sigma } \right),\hat x\left( {\tau
,\tilde \sigma } \right)} \right] = i\theta \delta \left( {\sigma -
\tilde \sigma
} \right),\\
\left[ {\hat t\left( {\tau ,\sigma } \right),\hat p_t\left( {\tau
,\tilde \sigma } \right)} \right] = i \delta \left( {\sigma  -
\tilde \sigma
} \right),\\
\left[ {\hat x\left( {\tau ,\sigma } \right),\hat p_x\left( {\tau
,\tilde \sigma } \right)} \right] = i\delta \left( {\sigma  - \tilde
\sigma
} \right),\\
\left[ {\hat \phi\left( {\tau ,\sigma } \right),\hat P_\phi\left(
{\tau ,\tilde \sigma } \right)} \right] = i\delta \left( {\sigma  -
\tilde \sigma } \right).
\end{gathered}
\end{equation}

 We turn now to the derivation of the explicit form for the modified first-class
constraints $\tilde{\mathcal H}_\perp$ and $\tilde{\mathcal H}_1$, by observing that the
formalism requires that their algebra should now close relative to
the Dirac-brackets. This can be achieved by further noting that
\begin{equation}
\{\tilde t, \tilde x\}^\ast=0,
\end{equation}
where
\be \label{tilded}
\tilde t=t+\frac{\theta}{2}p_x, \;\;\; \tilde x=x-\frac{\theta}{2}p_t.
\ee
This selection of the $\tilde t, \tilde x$, variables
is not unique, since there exist an infinite number of possible choices all of which
are related by canonical transformations that leave invariant the symplectic structure
(\ref{symft}). At the quantum level, however, only those theories which are related
by linear canonical transformations will be equivalent.
Now, taking into account that the Dirac-bracket algebra of the variables $(\tilde t, \tilde x, \phi, p_t, p_x, P_\phi)$
is the same as the Poisson algebra of $(t, x, \phi, p_t, p_x, P_\phi)$, it therefore
follows that by setting $\tilde{\mathcal H}_{\perp,1}(z^a) ={\mathcal H}_{\perp,1}(\tilde z^a)$ we
immediately have
\begin{eqnarray}\label{fcc3}
\{\tilde{\mathcal H}_\perp (\tau,\sigma) , \tilde{\mathcal H}_\perp (\tau,\sigma'\}^\ast &=&(\tilde{\mathcal H}_1 (\tau,\sigma)
+\tilde{\mathcal H}_1 (\tau,\sigma'))\delta'(\sigma-\sigma'),\nonumber\\
\{\tilde{\mathcal H}_1 (\tau,\sigma), \tilde{\mathcal H}_1 (\tau,\sigma')\}^\ast &=& (\tilde{\mathcal H}_1(\tau,\sigma)
+\tilde{\mathcal H}_1 (\tau,\sigma'))\delta'(\sigma-\sigma'),\\
\{{\mathcal H}_0 (\tau,\sigma), {\mathcal H}_1 (\tau,\sigma')\}^\ast
&=&(\tilde{\mathcal H}_\perp (\tau,\sigma) +\tilde{\mathcal
H}_\perp(\tau,\sigma'))\delta'(\sigma-\sigma'),\nonumber
\end{eqnarray}
with
\begin{equation}\label{ftcons2}
\begin{gathered}
  \tilde{\mathcal H} _\perp  =\frac{1}{2}( P_\phi ^2 + \phi^{'2}) + p_t (x-\frac{\theta}{2}p_t)'
  + p_x (t +\frac{\theta}{2}p_x)' +
  \left((x-\frac{\theta}{2}p_t)^{'2}   -(t +\frac{\theta}{2}p_x)^{'2} \right)V\left( \phi  \right)   \approx 0, \hfill \\
  \tilde{\mathcal H} _1  = p_x (x-\frac{\theta}{2}p_t)' + p_t (t +\frac{\theta}{2}p_x)' + P_\phi  \phi ' \approx 0. \hfill \\
\end{gathered}
\end{equation}

When quantizing, the constraints $\tilde{\mathcal H}_{\perp,1}$ are promoted to the rank of operators
satisfying the subsidiary conditions
\begin{equation}\label{qsub}
\begin{split}
\hat{\tilde {\mathcal H}}_{\perp} |\Psi\rangle &=0,\\
\hat{\tilde{\mathcal H}}_1 |\Psi\rangle &=0.
\end{split}
\end{equation}
Also for consistency we need that at the quantum level the additional condition
\begin{equation}\label{qcons1}
[ \hat{\tilde {\mathcal H}}_{\perp} ,\hat{\tilde {\mathcal H}}_{1}] |\Psi\rangle =0,
\end{equation}
be satisfied. This implies that the commutator of the first
class constraint operators has to be of the form
\begin{equation}\label{qcomm}
[ \hat{\tilde {\mathcal H}}_{\perp}(\tau,\sigma) ,\hat{\tilde {\mathcal H}}_{1}(\tau,\sigma')]
 =\hat c_\perp(\sigma,\sigma') \hat{\tilde{\mathcal H}}_\perp +\hat c_1 (\sigma,\sigma') \hat{\tilde{\mathcal H}}_1,
\end{equation}
where, in general, the ${\hat c}_{\perp,1}$ are functions of the field operators that need
to appear to the left of the $\hat{\tilde{\mathcal H}}_{\perp,1}$.
 This, in turn, involves
finding the operator ordering needed to achieve this requirement
in order to have an appropriate quantum theory. In the present case this does not
constitute an important issue, since ordering for the super-Hamiltonian is immaterial
and the difference in placing the momenta to the right or to the left of the coordinates in the
super-momentum leads to a term which in the basis $|t(\sigma), p_x (\sigma), \phi(\sigma)\rangle$
(see paragraph following Eq.(\ref{qcomm3}) below) is of the form
$\partial_{\sigma}\delta(\sigma-\sigma')|_{\sigma=\sigma'}
\Psi( t(\sigma),p_x(\sigma), \phi(x(\sigma), t(\sigma)),\tau)$ and which, because of the antisymmetry
of the delta function derivative, can be put equal to zero.
We therefore choose the following ordering for the $\hat{\tilde{\mathcal H}}_{\perp,1}$:
\begin{equation}\label{qftcons2}
\begin{gathered}
  \hat{\tilde{\mathcal H}}_\perp  =\frac{1}{2}\left(\hat{ P_\phi} ^2 + \hat\phi^{'2}\right) +
\hat {p}_t (\hat {x}-\frac{\theta}{2}\hat{p}_t)' + \hat{p}_x
(\hat{t} +\frac{\theta}{2}\hat{p}_x)' -
  \left( (\hat{t} +\frac{\theta}{2}\hat{p}_x)^{'2} -(\hat{x}-\frac{\theta}{2}\hat{p}_t)^{'2} \right)
V\left(\hat \phi  \right)   \approx 0, \hfill \\
\hat{\tilde{\mathcal H}} _1  = \hat{p}_x (\hat{x}-\frac{\theta}{2}\hat{p}_t)' + \hat{p}_t (\hat t +
\frac{\theta}{2}\hat p_x)' + \hat P_\phi  \hat\phi ' \approx 0. \hfill \\
\end{gathered}
\end{equation}
Making repeated use of the identity
\begin{equation}\label{ident}
f(\sigma')\delta'(\sigma-\sigma')= f'(\sigma)\delta(\sigma-\sigma')+f(\sigma)\delta'(\sigma-\sigma')
\end{equation}
in the evaluation of the commutator of these two operators, we get
\begin{equation}
\begin{split}
2\hat P_{\phi}(\sigma) \hat P_{\phi}(\sigma')\delta'(\sigma-\sigma')=(\hat P_{\phi}^2(\sigma) +
\hat P_{\phi}^2(\sigma'))\delta'(\sigma-\sigma'),\hspace{1.5in}\\
2\hat{\tilde x}'(\sigma)\hat{\tilde x}'(\sigma')\delta'(\sigma-\sigma')=
(\hat{\tilde x}'^2 (\sigma)+\hat{\tilde x}'^2 (\sigma'))\delta'(\sigma-\sigma'),\hspace{1.6in}\\
2\hat{\tilde t}'(\sigma)\hat{\tilde t}'(\sigma')\delta'(\sigma-\sigma')=
(\hat{\tilde t}'^2 (\sigma)+\hat{\tilde t}'^2 (\sigma'))\delta'(\sigma-\sigma'),\hspace{1.7in}\\
\left(\hat p_t(\sigma)\hat{\tilde x}'(\sigma')+\hat p_t(\sigma')\hat{\tilde x}'(\sigma)\right)\delta'(\sigma-\sigma')=
\left(\hat p_t(\sigma)\hat{\tilde x}'(\sigma)+\hat p_t(\sigma')\hat{\tilde x}'(\sigma')\right)\delta'(\sigma-\sigma'),\hspace{.1in}\\
\left(\hat p_x(\sigma)\hat{\tilde t}'(\sigma')+\hat p_x(\sigma')\hat{\tilde t}'(\sigma)\right)\delta'(\sigma-\sigma')=
\left(\hat p_x(\sigma)\hat{\tilde t}'(\sigma)+\hat p_x(\sigma')\hat{\tilde t}'(\sigma')\right)\delta'(\sigma-\sigma'),\hspace{.2in}\\
(\hat{\tilde x}'^2(\sigma)+\hat{\tilde t}'^2(\sigma))[V(\hat\phi(\sigma)), \hat P_\phi(\sigma')]\phi'(\sigma')=i
(\hat{\tilde x}'^2(\sigma')+\hat{\tilde t}'^2(\sigma'))\partial_\sigma V(\hat\phi(\sigma))\delta(\sigma-\sigma')\\
=i(\hat{\tilde x}'^2(\sigma')+\hat{\tilde t}'^2(\sigma'))\left(V(\hat\phi(\sigma'))-V(\hat\phi(\sigma))\right)\delta'(\sigma-\sigma').
\end{split}
\end{equation}
From these relations it follows that
\begin{equation}\label{qcomm3}
[\hat{\tilde{\mathcal H}}_\perp(\tau,\sigma), \hat{\tilde{\mathcal H}}_{1}(\tau,\sigma')]=
i \left(\hat{\tilde{\mathcal H}}_\perp (\tau,\sigma)+\hat{\tilde{\mathcal H}}_\perp(\tau,\sigma')\right)\delta'(\sigma-\sigma').
\end{equation}
Hence our choice (\ref{qftcons2}) is indeed of the form (\ref{qcomm}) and results in an appropriate Dirac
quantization of the theory.  In this parametrized quantization all the dynamics is hidden in the constraints although,
because of the noncommutativity of the coordinate field operators
$t(\tau, \sigma), x(\tau, \sigma)$, we can not construct configuration space state functionals
of the form $\Psi[t(\sigma), x(\sigma), \phi(\sigma),\tau] =\langle t(\sigma), x(\sigma), \phi(\sigma)|\Psi(\tau) \rangle$
with the usual interpretation of a probability amplitude that the scalar field $\phi$ have a definite
distribution $\phi(\sigma)$ on a curved spacelike hypersurface defined by $t=t(\sigma),\; x=x(\sigma)$
at time $\tau$. (Note that
in the Schr\"odinger picture the dynamical variables do not depend on $\tau$).
We can, however, construct state amplitudes from mixed momenta and reduced configuration space
eigenkets such as $|t(\sigma), p_x (\sigma), \phi(\sigma)\rangle$. In this basis $\hat x$ and $\hat p_t$
are represented by
\begin{eqnarray}
\hat x &=& i\left(\frac{\delta}{\delta p_x(\sigma)} -\theta\frac{\delta}{\delta t(\sigma)}\right),\\
\hat p_t &=& -i\frac{\delta}{\delta t(\sigma)},
\end{eqnarray}
so that from (\ref{qftcons2}) we get:
\be\label{ncschro1}
\begin{split}
\left(\frac{\theta}{2}\frac{\partial}{\partial
\sigma}\frac{\delta^2}{\delta t(\sigma) \delta t(\sigma)}
-\frac{\partial}{\partial \sigma}\frac{\delta}{\delta
t(\sigma)}\frac{\delta}{\delta p_x(\sigma)}
\right)\Psi[t(\sigma), p_x (\sigma), \phi(\sigma),\tau]=\hspace{2in}\\
\left[\frac{1}{2}\left(-\frac{\delta^2}{\delta\phi(\sigma)\delta\phi(\sigma)}+\phi'^2
\right) +p_x (t' +\frac{\theta}{2} p'_x) -\left((t' +
\frac{\theta}{2} p'_x)^2 +\frac{\partial^2}{\partial\sigma^2}
\left(\frac{\delta}{\delta p_x(\sigma)}-\frac{\theta}{2}
\frac{\delta}{\delta t(\sigma)}\right)^2
  \right)V(\phi)\right]\Psi,
\end{split}
\ee and \be \label{wd2}\left[p_x \frac{\partial}{\partial
\sigma}\left(\frac{\delta}{\delta p_x(\sigma)}- \frac{\theta}{2}
\frac{\delta}{\delta t(\sigma)} \right) -
\left(t'-\frac{\theta}{2}p'_x\right)\frac{\delta}{\delta t(\sigma)}
- \phi' \frac{\delta}{\delta \phi}\right]\Psi[t(\sigma), p_x
(\sigma), \phi(\sigma)]=0. \ee Thus, introducing noncommutativity by
parametrizing the action in the Dirac first quantization of the
scalar field scheme leads us necessarily to the above twofold
infinity of coupled equations. The equations (\ref{ncschro1}) and
(\ref{wd2}) are the analogous of the Wheeler-De Witt equations for
our noncommutative scalar field, and they can not be reduced to a
Schr\"odinger-like equation as in the commutative case, because here
we can not solve explicitly the super-Hamiltonian and super-momentum
constraints for the momenta $p_t$ and $p_x$. It is not our objective
here to investigate this system any further or the issue of second
quantization. We shall consider instead in the following section the
deformed symmetries which result from the deformed constraints of
the theory, which in turn result from the space-time
noncommutativity, and derive a general anzatz for constructing these
deformed symmetries for any field theory.

\section{spacetime noncommutativity and deformed symmetries}

We have seen that the Dirac-bracket algebra (\ref{fcc3}) together
with (\ref{ftcons2}) provides an algorithm for constructing the
deformed gauge symmetries associated with the reparametrization
invariance of the action (\ref{ftact}), where a symplectic structure
was introduced in order to allow for the appearance of spacetime
noncommutativity when applying Dirac's procedure for canonical
quantization to the original action. In fact, making use of
(\ref{comm}) one can show that \be\label{ident3}
\{t^{n}(\tau,\sigma) ,x^{m}(\tau,\sigma') \}^{\ast} = nm\theta
t^{n-1}(\tau,\sigma) x^{m-1} (\tau,\sigma')\delta(\sigma -\sigma').
\ee On the other hand, evaluating the Moyal product $(x^{\mu})^{n}
\star_{\theta} (x^{\nu})^{m} $ with the bidifferential
\be\label{star} \star_{\theta}:=
\exp\left[\frac{i}{2}\theta^{\mu\nu}\int
d\sigma''\frac{\overleftarrow\delta} {\delta
x^{\mu}(\tau,\sigma'')}\frac{\overrightarrow\delta}{\delta
x^{\nu}(\tau,\sigma'')}\right] , \ee and comparing with
(\ref{ident3}), we have that \be\label{iso} \{t^{n}(\tau,\sigma)
,x^{m}(\tau,\sigma') \}^{\ast} \cong
[t^{n}(\tau,\sigma),x^{m}(\tau,\sigma')]_{ \star_{\theta}}:=
 t^{n}(\tau,\sigma)\star_{\theta} x^{m}(\tau,\sigma') - x^{m}(\tau,\sigma') \star_{\theta} t^{n}(\tau,\sigma).
\ee
More generally, for Dirac- brackets of arbitrary $A(\tau,\sigma)$, $B(\tau,\sigma)$ functionals
of $t(\tau,\sigma),\; p_t (\tau,\sigma),\;
 x(\tau,\sigma),\\p_x (\tau,\sigma),\;\phi(\tau,\sigma)$ and $P_\phi (\tau,\sigma)$  we get
\be\label{morf3} \{A(\tau,\sigma), B(\tau,\sigma')\}^\ast \cong
[A(\tau,\sigma),B(\tau,\sigma')]_{\star_\theta}, \ee after
identifying the momenta in the left side of the above equation with
their corresponding differential operators on the right side. We
thus have a morphism from the Poisson-Dirac algebra of functionals
of $t, x, \phi, p_t , p_x\; \text {and}\; P_\phi $, to the algebra
of differential operators obtained from these functionals (after
making the maps $p_t \mapsto -i\delta/\delta_t , \;\; p_x \mapsto
-i\delta/\delta_x\;\;P_\phi \mapsto -i\delta/\delta_\phi $)
 with multiplication given by the $\star_\theta$-product  commutator. As a parenthetical remark
we find it interesting to recall here that in the process of
reparametrization the space-time parameters of the original action
were elevated to the rank of dynamical variables and, as we have
shown elsewhere \cite{ros}, when considering quantum mechanical
deformations from the point of view of the
Weyl-Wigner-Groenewold-Moyal formalism, the multiplication of
elements of the algebra of functions of the space-time dynamical
variables had to be modified precisely
with the $\star$-operator (\ref{star}).\\

Applying now the above described algebra morphism to (\ref{fcc3}) results in
\begin{eqnarray}\label{fcc4}
\left[\tilde{\mathcal H}^{\star}_\perp (\tau,\sigma) , \tilde{\mathcal H}^{\star}_\perp (\tau,\sigma')\right]_{\star\theta}
&=&\left(\tilde{\mathcal H}^{\star}_1 (\tau,\sigma)
+\tilde{\mathcal H}^{\star}_1 (\tau,\sigma')\right)\delta'(\sigma-\sigma'),\nonumber\\
\left[\tilde{\mathcal H}^{\star}_1 (\tau,\sigma) , \tilde{\mathcal H}^{\star}_1 (\tau,\sigma')\right]_{\star\theta}
&=&\left(\tilde{\mathcal H}^{\star}_1 (\tau,\sigma)
+\tilde{\mathcal H}^{\star}_1 (\tau,\sigma')\right)\delta'(\sigma-\sigma'),\\
\left[\tilde{\mathcal H}^{\star}_\perp (\tau,\sigma), \tilde{\mathcal H}^{\star}_1 (\tau,\sigma')\right]_{\star\theta}
 &=&\left(\tilde{\mathcal H}^{\star}_\perp (\tau,\sigma)
+\tilde{\mathcal
H}^{\star}_\perp(\tau,\sigma')\right)\delta'(\sigma-\sigma').\nonumber
\end{eqnarray}
Here the notation $\tilde{\mathcal H}^{\star}_{\perp,1}$ stands for the differential operators
\be\label{derivation}
\tilde{\mathcal H}^{\star}_{\perp,1} (\tau,\sigma) := {\mathcal H}_{\perp,1}(\tau,\sigma)
\exp\left[-\frac{i}{2}\theta^{\mu\nu}\int d\sigma''\frac{\overleftarrow\delta}
{\delta x^{\mu}(\tau,\sigma'')}\frac{\overrightarrow\delta}{\delta x^{\nu}(\tau,\sigma'')}\right] ,
\ee
and their algebra multiplication $\mu_{\theta}$ is given by
\be\label{starmult}
\mu_{\theta}(\tilde{\mathcal H}^{\star}_i \otimes \tilde{\mathcal H}^{\star}_j)=
 \tilde{\mathcal H}^{\star}_i \star \tilde{\mathcal H}^{\star}_j,\;\;\; i,j=\perp, 1 .
\ee
Note that from (\ref{derivation}) it follows that
\begin{equation}\label{noncom}
\left[\tilde{\mathcal H}^{\star}_i (\tau,\sigma) , \tilde{\mathcal H}^{\star}_j  (\tau,\sigma')\right]_{\star\theta}=
\left[{\mathcal H}_i (\tau,\sigma), {\mathcal H}_j (\tau,\sigma')\right] e^{\left[-\frac{i}{2}\theta^{\mu\nu}\int d\sigma''\frac{\overleftarrow\delta}
{\delta x^{\mu}(\tau,\sigma'')}\frac{\overrightarrow\delta}{\delta x^{\nu}(\tau,\sigma'')}\right]},\;\;\;i,j=\perp,1
\end{equation}
and substituting (\ref{derivation}) and (\ref{noncom}) into (\ref{fcc4}) we get
\begin{eqnarray}\label{fcc5}
\left[{\mathcal H}_\perp (\tau,\sigma) , {\mathcal H}_\perp (\tau,\sigma')\right] &=&\left({\mathcal H}_1 (\tau,\sigma)
+{\mathcal H}_1 (\tau,\sigma')\right)\delta'(\sigma-\sigma'),\nonumber\\
\left[{\mathcal H}_1 (\tau,\sigma) , {\mathcal H}_1 (\tau,\sigma')\right] &=&\left({\mathcal H}_1 (\tau,\sigma)
+{\mathcal H}_1 (\tau,\sigma')\right)\delta'(\sigma-\sigma'),\\
\left[{\mathcal H}_\perp (\tau,\sigma),{\mathcal H}_1
(\tau,\sigma')\right]&=&\left({\mathcal H}_\perp (\tau,\sigma)
+{\mathcal
H}_\perp(\tau,\sigma')\right)\delta'(\sigma-\sigma'),\nonumber
\end{eqnarray}
which is the algebra  of differential operator generators isomorphic to the non-deformed algebra (\ref{fcc2}).\\

Furthermore, since by (\ref{ftcons})
\begin{equation}\label{deriv2}
\begin{gathered}
\{\phi, {\mathcal H}_\perp \}=\frac{(x'^{2} -t'^{2})}{\sqrt {-g}}
\dot\phi+
\frac{(t'\dot t-x'\dot x)}{\sqrt {-g}} \phi',\\
\{\phi, {\mathcal H}_1 \}=\phi',
\end{gathered}
\end{equation}
the generators
${\mathcal H}_i$ of  (\ref{fcc2}) - the Virasoro algebra ${\mathcal V}$ - can be viewed  as derivations acting on elements
$\phi(t(\tau,\sigma),x(\tau,\sigma))$ of the algebra of functions $\mathcal A$, with point multiplication $\mu$.
That is,
\begin{equation}\label{deriv3}
\begin{split}
\{\phi, {\mathcal H}_\perp\}\cong& \hat{\mathcal H}_\perp
\triangleright \phi=\left(\frac{(x'^{2} -t'^{2})}{\sqrt {-g}}
\partial_\tau+
\frac{(t'\dot t-x'\dot x)}{\sqrt {-g}} \partial_\sigma\right) \triangleright\phi\\
\{\phi, \hat{\mathcal H}_1 \}\cong& \hat{\mathcal H}_1
\triangleright \phi =\partial_\sigma \triangleright \phi,
\end{split}
\end{equation}
In addition, since $\hat{\mathcal H}_i \in \hat{\mathcal V}$ is a
(infinite dimensional) Lie algebra, its universal envelope
$U(\hat{\mathcal V})$ can be given the structure of a Hopf algebra
with coproduct \be\label{coprod} \Delta(\hat{\mathcal
H}_i)=\hat{\mathcal H}_i \otimes 1 +1\otimes \hat{\mathcal H}_i
,\;\;\;i=\perp,1 \ee and antipode \be\label{antip} S(\hat{\mathcal
H}_i)=-\hat{\mathcal H}_i,\;\;\;i=\perp,1 , \ee so $\mathcal A$ is a
left module-algebra over $U(\hat{\mathcal V})$. In parallel, for the
symplectic structure (\ref{symft}) we have the algebra
$\hat{\mathcal V}^{\star}$ of derivation operators
$\hat{\tilde{\mathcal H}}^{\star}_i$, defined in analogy to
(\ref{derivation}) by \be\label{derivation2} \hat{\tilde{\mathcal
H}}^{\star}_{\perp,1} (\tau,\sigma) := \hat{\mathcal
H}_{\perp,1}(\tau,\sigma) \exp\left[-\frac{i}{2}\theta^{\mu\nu}\int
d\sigma''\frac{\overleftarrow\delta} {\delta
x^{\mu}(\tau,\sigma'')}\frac{\overrightarrow\delta}{\delta
x^{\nu}(\tau,\sigma'')}\right] , \ee with multiplication
$\mu_\theta$ generated by (\ref{morf3}), and the corresponding left
module algebra ${\mathcal A}_\theta $ over $U(\hat{\mathcal
V}^\star)$, whose elements are now functions $\phi
(t(\tau,\sigma),x(\tau,\sigma))$ with multiplication
$\mu_\theta$ inherited from (\ref{iso}).\\
From (\ref{derivation2}) it immediately follows that
\be\label{modact} \hat{\tilde{\mathcal H}}^{\star}_i\star_{\theta}
\phi(t,x) =\hat{\mathcal H}_i \triangleright \phi(t,x), \ee so the
action of elements of the twisted algebra $\hat{\mathcal V}^\star$
on elements of  ${\mathcal A}_\theta $ is equal to the action of the
corresponding elements of the untwisted algebra on the corresponding
elements of the ordinary algebra $\mathcal A$ of functions of
commuting variables. Thus the morphism from $\hat{\mathcal V}$ to
$\hat{\mathcal V^\star}$ by \be\label{morf2}
  \hat{\mathcal H}_i\mapsto \hat{\tilde{\mathcal H}}^{\star}_i
 \ee
induces the morphism from ${\mathcal A} $ to ${\mathcal A}_\theta $ by
\be\label{morf4}
\mu(f(t,x)\otimes g(t,x))\mapsto \mu_\theta (f(t,x)\otimes g(t,x)).
\ee

Let us next consider the symmetries associated with the canonical
transformation \be\label{generator} H_{\tau}[\xi]= \int d\sigma
\left (\xi^0(\tau,\sigma) {\mathcal H}_\perp(\tau,\sigma)+
\xi^1(\tau,\sigma) {\mathcal H}_1(\tau,\sigma) \right), \ee in order
to make contact with some related  results appearing in the
literature. We thus have
\begin{eqnarray}\label{cantrans}
 \delta\phi&=& \{\phi,H_{\tau}[\xi]\}\cong \hat H_{\tau}[\xi]\triangleright\phi=\xi^0 \frac{(x'^{2} -t'^{2})}{\sqrt {-g}} \dot\phi+
\left(\xi^0 \frac{(t'\dot t-x'\dot x)}{\sqrt {-g}} +\xi^1 \right)\phi' \nonumber\\
\delta t&=&\{t,H_{\tau}[\xi]\}=\xi^0 x' +\xi^1 t' \\
\delta x&=& \{x,H_{\tau}[\xi]\}= \xi^0 t' +\xi^1 x'.\nonumber
\end{eqnarray}

On the other hand, it is evident that the Lagrangian in (\ref{ftact}) is invariant
under the infinitesimal general coordinate transformations
\begin{equation}\label{gct}
\begin{split}
\tau &\rightarrow \tau +\rho^0 (\tau,\sigma)\\
\sigma &\rightarrow  \sigma +\rho^1 (\tau,\sigma),
\end{split}
\end{equation}
from where it follows that
\begin{eqnarray}\label{cantrans2}
 \delta_\rho\phi&=& -\rho^0 \partial_{\tau} \phi -\rho^1 \partial_{\sigma} \phi,\nonumber\\
\delta_\rho t&=& -\rho^0 \partial_{\tau} t -\rho^1 \partial_{\sigma} t,\\
\delta_\rho x&=& -\rho^0 \partial_{\tau} x -\rho^1 \partial_{\sigma} x.\nonumber
\end{eqnarray}
We can relate the generator (\ref{generator}) to the diffeomorphism (\ref{cantrans2})
by equating the last two equations in (\ref{cantrans}) to the last two equations in (\ref{cantrans2})
and solving for $\xi^0$ and $\xi^1$. We thus get
\begin{equation}\label{algmorf}
\begin{split}
\xi^0 &= \frac{(t' \dot x -x' \dot t)}{(x'^2 -t'^2)}\rho^0,\\
\xi^1 &=  \frac{(t' \dot t -x' \dot x)}{(x'^2 -t'^2)}\rho^0 -\rho^1.
\end{split}
\end{equation}
The consistency of this solution can be checked by substituting it
into the first equation in (\ref{cantrans}) and verifying that it
yields the first equation in (\ref{cantrans2}). Consequently
\be\label{difeo} \delta_{\rho} \phi=\hat
H_{\tau}[\xi(\rho)]\triangleright\phi\cong\{\phi,H_{\tau}[\xi(\rho)]\},
\ee with the components of $\xi(\rho)$ given by (\ref{algmorf}).
Hence \be\label{deriv31} \delta_\rho = \hat H_\tau [\xi(\rho)] =
-(\rho^{0} \dot t +\rho^{1} t')\partial_t -(\rho^{0} \dot x
+\rho^{1} x')\partial_x =-(\rho^t \partial_t +\rho^x \partial_x ),
\ee
where we have re-expressed the vector field $\delta_\rho$ in terms of the spacetime basis components\\
$\{\rho^t :=(\rho^{0} \dot t +\rho^{1} t'),\;\; \rho^x :=(\rho^{0} \dot x +\rho^{1} x')\}$.\\
Applying now the derivation $\delta_\eta :=-\eta^t \partial_{t}-\eta^x \partial_{x}$ to (\ref{difeo}) and
subtracting from the result the expression with inverted order of the derivations we get
\begin{eqnarray}\label{difeo2}
[\delta_\eta , \delta_\rho ]\phi&\cong& \{\{\phi, H_\tau [\xi(\rho)]\}, H_\tau [\xi(\eta)]\} -
\{\{\phi, H_\tau [\xi(\eta)]\}, H_\tau [\xi(\rho)]\}=\{\{ H_\tau [\xi(\eta)],H_\tau [\xi(\rho)]\},\phi\}\nonumber\\
&=& -(-\eta^\lambda \partial_{\lambda} \rho^\mu +
 \rho^\lambda \partial_{\lambda} \eta^\mu ) \partial_{\mu} \phi =-(\eta\times\rho)^\mu \partial_{\mu} \phi
=\delta_{ \eta\times\rho} \;\phi,
\end{eqnarray}
after making use of the Jacobi identity.
We therefore have an homomorphism between the algebra of diffeomorphisms
in two-dimensions
\be\label{antihomo}
[\delta_\eta , \delta_\rho ]=\delta_{ \eta\times\rho}
\ee
and the Poisson algebra $\frak H$ generated by
\be\label{virared}
 \{ H_\tau [\xi(\eta)],H_\tau [\xi(\rho)]\}=H_\tau [\xi(\eta\times\rho)].
\ee
In going on to the noncommutative spacetime case, we proceed
according to our previously derived algorithm, {\it i.e.} we replace
the Poisson-brackets by Dirac-brackets and $t\rightarrow \tilde t$,
$x\rightarrow \tilde x$. Hence
we can now write
\be \label{derop}
{\frak H}\ni \hat H_{\tau}[\xi(\rho)]\mapsto {\hat H}^{\star}_{\tau}[\tilde\xi(\rho)]=\delta^\star_\rho=
\int d\sigma(\tilde\xi^0 \hat{\tilde{\mathcal H}}^{\star}_\perp +
\tilde\xi^1 \hat{\tilde{\mathcal H}}^{\star}_1) \in {\frak H}^\star,
\ee
and
\be\label{derop2}
\{\phi, H_{\tau}[\xi(\rho)]\}\cong \delta_\rho\triangleright \phi\mapsto
\delta^{\star}_{\rho} \star\phi(t(\tau,\sigma), x(\tau,\sigma));\;\;\;\phi\in {\mathcal A}_\theta.
\ee
 Note that equations (\ref{derop}) and (\ref{derop2}) provide an explicit expression
for the mapping $\delta_\rho \mapsto \delta^{\star}_{\rho} $, such that
(\ref{virared}) becomes
\be\label{virared2}
\left[\delta^{\star}_{\rho}, \delta^{\star}_{\eta}\right]_{\star\theta}=\delta^\star_{\eta\times\rho},
\ee
and
\be\label{virared3}
\delta^{\star}_{\rho} \star(f\star g) =\delta_\rho (f\star g).
\ee
We can now compare some of our results with those obtained in \cite{wess}.
Thus, we have that our equation (\ref{derivation2}) for the twisted
derivations $\hat{\tilde{\mathcal H}}^{\star}_i$ corresponds to equation (3.26)
in \cite{wess}, while the algebra (\ref{virared2}) and the derivation
$\delta^{\star}_{\rho}$ correspond to equations
(5.3) and (5.4) there.
Note also that since the universal envelope
$U({\frak H}^{\star})$ in our formalism can be given the structure of a Hopf algebra,
we can obtain an explicit expression for the coproduct by making use of the duality
between product and coproduct, followed by the application of equations
(\ref{virared3}) and (\ref{derivation}). Thus we have
\be\label{defhopf}
\begin{split}
\mu_{\theta}\circ\Delta (\delta^{\star}_{\rho})(f\otimes g)=
\delta^{\star}_{\rho} \star(f\star g)=\delta_{\rho} (f\star g)=\\
\mu(\delta_{\rho} \otimes 1+1\otimes \delta_{\rho})
(e^{\frac{i}{2}\theta^{\mu\nu} \partial_{\mu}\otimes \partial_{\nu}} f\otimes g)=\\
\mu_{\theta}\circ\left[(\delta^{\star}_{\rho} \otimes 1 +1\otimes\delta^{\star}_{\rho})
 e^{\frac{i}{2}\theta^{\mu\nu} \partial_{\mu}\otimes \partial_{\nu}}(f\otimes g)\right]=\\
\mu_{\theta}\circ \left [e^{-\frac{i}{2}\theta^{\mu\nu}
\partial_{\mu}\otimes \partial_{\nu}}(\delta_{\rho}\otimes 1 + 1\otimes\delta_{\rho})
e^{\frac{i}{2}\theta^{\mu\nu} \partial_{\mu}\otimes \partial_{\nu}}\right](f\otimes g).
\end{split}
\ee
This result also compares with the Leibnitz rule given by equation (5.9) in \cite{wess}.
Further note that if we let
${\mathcal F}=e^{-\frac{i}{2}\theta^{\mu\nu}\partial_{\mu}\otimes \partial_{\nu}}\in
U({\frak H})\otimes U({\frak H})$, and define
$f\star g=\mu_\theta (f\otimes g):=\mu({\mathcal F}^{-1}\triangleright (f\otimes g))$,
we then have
\begin{eqnarray}\label{drintwist}
\delta_{\rho}(f\star g)&=&\delta_{\rho}\triangleright
\mu({\mathcal F}^{-1}\triangleright (f\otimes g))=
\mu[(\Delta \delta_{\rho}){\mathcal F}^{-1}\triangleright (f\otimes g)]\nonumber\\
&=& \mu {\mathcal F}^{-1}[({\mathcal F}(\Delta \delta_{\rho}) {\mathcal F}^{-1})(f\otimes g)]\\
&=&\mu_\theta [({\mathcal F}(\Delta \delta_{\rho}) {\mathcal F}^{-1})(f\otimes g)].\nonumber
\end{eqnarray}
Thus, the undeformed coproduct of the symmetry Hopf algebra $U({\frak H})$ is
related to the Drinfeld twist $\Delta^{\mathcal F}$ by the inner endomorphism
$\Delta^{\mathcal F}\delta_\rho := ({\mathcal F}(\Delta \delta_{\rho}) {\mathcal F}^{-1})$
and, by (\ref{drintwist}), it preserves the covariance:
\begin{eqnarray}
\delta_\rho \triangleright (f\cdot g)&=&\mu\circ [\Delta (\delta_\rho)(f\otimes g)]=
(\delta_{\rho (1)} \triangleright f)\cdot (\delta_{\rho (2)} \triangleright g)\nonumber\\
&\stackrel{\theta}{\rightarrow}& \delta^\star_\rho \triangleright (f\star g)=
(\delta^\star_{\rho (1)}\: \triangleright f)\star (\delta^\star_{\rho (2)})\: \triangleright g),
\end{eqnarray}
where we have used the Sweedler notation for the coproduct. Consequently, the twisting of the
coproduct is tied to the deformation $\mu\rightarrow \mu_\theta$ of the product when the last
one is defined by
\be\label{prodef}
f\star g := ({\mathcal F}^{-1}_{(1)}\:\triangleright f) ({\mathcal F}^{-1}_{(2)}\:\triangleright g).
\ee

A more extensive discussion of the application of some of these
algebras to the construction of a deformed differential geometry for
gravity theories may be found also in \cite{wess} as well as other
works cited therein.

If we now assume that the coefficients of the vector fields $\delta_\xi$ are linear in the spacetime
variables, then the generators $\delta_{\rho}$ in (\ref{defhopf}) become the infinitesimal
generators of the Poincar\'e transformations, and the coproduct defined in this equation reduces to the
twisted coproduct considered by {\it e.g.} \cite{chaichian}.

We would like to stress, however, that while all the above mentioned papers, as well
as a large number of others appearing in the literature, start
from equating spacetime noncommutativity with the noncommutativity of the parameters of
the functions denoting classical fields, and deforming the algebra of these fields
via the Moyal $\star$-product (with this anzatz originating in a basically heuristic
fashion from some results in string theory), none of the algebras
 $\hat {\mathcal V}^{\star}, {\frak H}^{\star}$ and ${\mathcal A}_{\theta} $ in our approach
are assumed a {\it priori}. On the contrary, they appear naturally,
as does the spacetime noncommutativity, as a consequence of
implementing Dirac's canonical quantization formalism for
constrained systems with an arbitrary symplectic structure. Note, in
particular, that in our formalism the space-time variables are
dynamical, as would be expected when viewing quantum mechanics as a
minisuperspace of field theory, and their noncommutativity results
from the quantization of their Dirac-brackets. The deformation of
the module-algebra $\mathcal A$ - in which the fields originally
lived - to ${\mathcal A}_\theta\ni\phi$, so that by (\ref{morf2})
and (\ref{morf4}) functions of the field multiply according to
$\mu_\theta$ is, in our formalism, again a consequence of the
spacetime noncommutativity resulting from the quantization of the
Dirac-brackets, and the concomitant deformation of the constraints
associated with the symmetries of the field Lagrangian.

Finally, it should be obvious by mere observation of the notation
already introduced, how our algorithm can be readily extended to
higher dimensional noncommutative space-times with constant
parameters of noncommutativity. Thus, the commutator relations for
the spacetime coordinate fields at equal times will now be given by
\begin{equation}\label{comrftd}
    [x^\mu(\tau,\boldsymbol\sigma), x^\nu(\tau,\boldsymbol\sigma')]=i
    \theta^{\mu\nu} \delta^{D}(\boldsymbol\sigma -\boldsymbol\sigma'),
\end{equation}
where $\theta^{\mu\nu}=const. $ As in the bi-dimensional case, we
can also introduce a new set of commuting coordinate fields defined
by
\begin{equation}\label{tildesd}
\tilde x^\mu(\boldsymbol\sigma)= x^\mu (\boldsymbol\sigma) + \frac{\theta^{\mu\nu}}{2}
p_\nu(\boldsymbol\sigma),
\end{equation}
from which new constraints can be constructed having the form
\begin{equation}\label{consdp12}
\begin{split}
\tilde {\mathcal H}_\perp &=\frac{1}{2}\left( P_\phi ^2 +\tilde
\gamma \tilde \gamma^{ij}\partial_{\sigma^i}\phi
\partial_{\sigma^j} \phi \right) + \sqrt{-\tilde\gamma}\;\tilde n^\nu p_\nu -
  \tilde \gamma V\left( \phi  \right)  \approx 0, \\
\tilde {\mathcal H}_i &= P_\phi \partial_{\sigma^i} \phi + p_\mu
\partial_{\sigma^i} \tilde x^\mu.
\end{split}
\end{equation}
Making use the algebra morphism discussed at the beginning of this
section we then arrive at the quantum algebra
\begin{eqnarray}\label{fcc4d1}
\left[\tilde{\mathcal H}^{\star}_\perp (\tau,\boldsymbol\sigma) , \tilde{\mathcal H}^{\star}_\perp
(\tau,{\boldsymbol\sigma}')\right]_{\star\theta} &=&\sum_{i=1}^{D}\left(\tilde{\mathcal H}^{\star}_i
(\tau,\boldsymbol\sigma)
+\tilde{\mathcal H}^{\star}_i (\tau,{\boldsymbol\sigma}')\right)\partial_{\sigma^i}\delta(\boldsymbol\sigma -{\boldsymbol\sigma}'),\nonumber\\
\left[\tilde{\mathcal H}^{\star}_i (\tau,\boldsymbol\sigma) , \tilde{\mathcal H}^{\star}_j
(\tau,{\boldsymbol\sigma}')\right]_{\star\theta} &=&\left(\tilde{\mathcal H}^{\star}_i
(\tau,\boldsymbol\sigma)\partial_{\sigma^j}\delta(\boldsymbol\sigma-{\boldsymbol\sigma}')
+\tilde{\mathcal H}^{\star}_j (\tau,\boldsymbol\sigma')\partial_{\sigma^i}
\delta(\boldsymbol\sigma-\boldsymbol\sigma')\right),\\
\left[\tilde{\mathcal H}^{\star}_\perp (\tau,\boldsymbol\sigma), \tilde{\mathcal H}^{\star}_i
(\tau,{\boldsymbol\sigma}')\right]_{\star\theta} &=&\left(\tilde{\mathcal H}^{\star}_\perp
(\tau,\boldsymbol\sigma)
+\tilde{\mathcal H}^{\star}_\perp(\tau,{\boldsymbol\sigma}')\right)
\partial_{\sigma^i}\delta(\boldsymbol\sigma-{\boldsymbol\sigma}').\nonumber
\end{eqnarray}
With the constraints (\ref{consdp12}) it is possible to construct a
quantum theory in the Schr\"odinger representation analogous to
(\ref{ncschro1}) and (\ref{wd2}). As in that case, however, since these constraints
are no longer linear and algebraic in the momenta (they contain
mixed products of the $p_\mu$'s and their derivatives), it is not
possible to solve explicitly for the spacial momenta in order to
construct a Schr\"odinger type equation. Nonetheless, it is still
possible to show that the action in the reduced configuration space
is in agreement with the usually proposed noncommutative field
theory for a scalar field.\\

As for the generalization to $(D+1)$-Minkowski spacetime of the
symmetries and twisted symmetries elaborated above for the $D=1$
case, the results follow through directly by replacing $\xi^0 ,\xi^1$
by
\begin{eqnarray}
\xi^0 &=&- \frac{1}{g^{00}\sqrt {-g}}\rho^0,\\
\xi^i &=&\frac{g^{0i}}{g^{00}}\rho^0 -\rho^i.
\end{eqnarray}
These expressions can be inferred immediately from (\ref{algmorf}).

\section{Concluding remarks}

We have shown in this paper how, by considering a parametrized
field theory, it is possible to introduce spacetime noncommutativity
from first principles. We have accomplished this by resorting to an
extended phase-space, leading to second class constraints which, in
order to remove them according to the Dirac quantization procedure, lead
in turn to Dirac-brackets. The latter
then result in a deformed symplectic structure for the spacetime coordinates and
corresponding canonical momenta, which yield the desired noncommutativity.

An important characteristic of our formulation is the automatic
deformation of the symmetry generators when the symplectic structure is deformed.
Such a deformation being imposed by the consistency conditions on the constraints (see
discussion in subsection 2.2),
which have as a result that the algebra of the deformed constraints is maintained
in the noncommutative case. This
provides us then with a straightforward algorithm for constructing the
Drinfeld twist of the Hopf algebras that one can associate with the reparametrization
symmetry groups. In addition, our formalism can be readily extended to
spacetimes of any dimensions and to the consideration of different
possible types of deformed products, of which the Moyal product is just a particular case.
Thus the formalism here described may turn out to be also useful for
achieving a better understanding of twisted symmetries in Yang-Mills field theories,
since in this case, in addition to the constraints associated with the
reparametrization, we will also have the constraints associated with invariance
under the gauge transformations
$$A_\mu(x)\rightarrow U(x)A_\mu(x)U^{-1}(x) + iU(x)\partial_\mu U^{-1}(x),$$
so the full set must then be analyzed in order to see how it is to be twisted
when noncommutativity is introduced.

\end{document}